\documentclass[onecolumn,preprint,showpacs,preprintnumbers,amsmath,amssymb,floatfix]{revtex4-1}
\usepackage{color}
\usepackage{graphicx}
\usepackage{float}
\usepackage{subfigure}
\begin{document}

\title{Excess noise in scanning tunneling microscope-style break junctions at room temperature}

\author{Ruoyu Chen$^{1}$, Patrick J. Wheeler$^{1}$, D.~Natelson$^{1, 2}$}

\affiliation{$^{1}$ Department of Physics and Astronomy, Rice University, 6100 Main St., Houston, TX 77005}
\affiliation{$^{2}$ Department of Electrical and Computer Engineering, Rice University, 6100 Main St,.Houston, TX 77005}

\date{\today}


\begin{abstract} 
Current noise in nanoscale systems provides additional information beyond the electronic conductance.  We report measurements at room temperature of the nonequilibrium ``excess'' noise in ensembles of atomic-scale gold junctions repeatedly formed and broken between a tip and a film, as a function of bias conditions.  We observe suppression of the noise near conductances associated with conductance quantization in such junctions, as expected from the finite temperature theory of shot noise in the limit of few quantum channels.   In higher conductance junctions, the Fano factor of the noise approaches 1/3 the value seen in the low conductance tunneling limit, consistent with theoretical expectations for the approach to the diffusive regime.  At conductance values where the shot noise is comparatively suppressed, there is a residual contribution to the noise that scales quadratically with the applied bias, likely due to a flicker noise/conductance fluctuation mechanism.
\end{abstract}

\maketitle

Shot noise, first discussed by Schottky in 1918\cite{Schottky:1918}, comprises fluctuations in the steady-state, nonequilibrium current that originate from the discreteness of the electron charge.  This is ``excess'' noise in addition to the Johnson-Nyquist\cite{Johnson:1928,Nyquist:1928} current fluctuations that are present at equilibrium in the absence of an applied bias current.   Shot noise tends to be suppressed in macroscopic structures at finite temperatures due to electron-phonon interactions.  In many mesoscopic systems small compared to the inelastic scattering length for the electrons, shot noise survives and is strongly related to the quantum nature of transport\cite{Blanter:2000}.  Many measurements have been performed in this regime on various devices in past two decades, including quantum point contacts\cite{Reznikov:1995,Kumar:1996}, diffusive metal conductors\cite{Liefrink:1994,Henny:1999},  break junctions\cite{vandenBrom:1999,Djukic:2006}, and quantum Hall systems\cite{Saminadayar:1997,dePicciotto:1997}.  Most of these experiments are conducted at cryogenic temperatures to avoid thermal smearing of the noise, though shot noise measurements are possible at room temperature in sufficiently nanoscale structures\cite{Wheeler:2010}.

The classical Schottky shot noise power in the current is~$S_{I}=2eI$~, where ~$S_{I} $~ is the spectral density of shot noise, expressed as the mean squared variation in the current ~$\langle \Delta I^{2} \rangle$~ per unit frequency.  Here $e$ is the magnitude of the electron charge, and $I$ is the average DC bias current.  This expression is derived assuming the arrival of charge carriers is Poisson distributed, with each electron unaffected by the arrival of a previous electron.  Deviations from Poissonian statistics may alter the noise, and these changes are usually expressed in terms of a Fano factor, $F$, such that the measured noise $S_{I}=2eI \times F$.  Values of $F\ne 1$ provide clues about the possible effects of interactions and underlying transport processes.  The shot noise of mesoscopic conductors at zero temperature is expressed\cite{Blanter:2000} in terms of quantum channels:
\begin{equation}
S_{I}=2eVG_{0}\sum_{i}^N{\tau_i(1-\tau_i)}
\label{eq:zeroT}
\end{equation}
where ~$G_{0}=2e^2/h$~ is the quantum of conductance, $V$ is the bias voltage across the junction, and ~$\tau_i$~ is the transmission probability of the $i$th quantum channel.  
Combining with the Landauer formula of ~$G=G_0\sum^{N}_{i}\tau_i$~, the Fano factor at zero temperature is:
\begin{equation}
F=\frac{\sum_{i}^N{\tau_i(1-\tau_i)}}{\sum_i^N{\tau_i}}
\label{eq:Fanodef}
\end{equation}
The Fano factor carries extra information about the transmission probabilities that a conductance measurement alone cannot provide.  At nonzero temperature (though assuming that energy is not exchanged between the charge carriers and other degrees of freedom such as phonons), the situation is more complex, as the 
thermal Johnson-Nyquist noise and shot noise are not readily separable. The total current noise will be:
\begin{equation}
S_{I}=G_0[4k_{\mathrm{B}}T\sum_i^N{\tau_i^{2}}+2eV \coth(\frac{eV}{2k_{\mathrm{B}}T})\sum_{i}^N{\tau_i(1-\tau_i)}]
\label{eq:finiteT}
\end{equation}
$k_{\mathrm{B}}$~is the Boltzmann constant.  In the equilibrium limit $V=0$, both terms in this formula will survive and contribute to Johnson-Nyquist thermal noise of ~$4k_{\mathrm{B}}TG$.   In the zero temperature limit, the total noise power will reduce to Eq.~(\ref{eq:zeroT}).   Temperature manifests itself through the smearing of the Fermi-Dirac distribution of the electrons.

From these equations it is clear that fully transmitting channels ($\tau_{i} \rightarrow 1$) do not contribute to the shot noise.   This leads to a relative suppression of the noise in nanoscale systems when the conductance is largely from such open channels, as in semiconductor point contacts exhibiting quantized conductance\cite{Reznikov:1995}, and in metal point contacts\cite{vandenBrom:1999}.  In the few channel limit, the conductance combined with the noise allow the determination of the number of channels and their transmission probabilities\cite{Djukic:2006}.   In the many-channel limit of diffusive conductors, random matrix theory has provided valuable insights, and the Fano factor is expected to approach an average value between $1/3$ and $\sqrt{3}/4$\cite{Beenakker:1992,Shimizu:1992,Nagaev:1992,deJong:1995} depending on bias conditions and sample geometry.  These predictions have been confirmed in the low temperature limit\cite{Liefrink:1994,Steinbach:1996,Schoelkopf:1997,Henny:1999}.  

Inelastic processes such as the excitation of local vibrational modes are predicted to alter $F$ as the bias voltage exceeds the energy scale of such excitations\cite{Mitra:2004,Koch:2005,Koch:2006}.  Such effects have been observed at low temperatures in nanotubes\cite{Leturcq:2009}, bilayer graphene\cite{Fay:2011}, and very recently in atomic-scale Au junctions\cite{Kumar:2012}, though the particular changes in $F$ depend in detail on channel transmission.  One motivation of this work is the need to perform experimental comparisons with Eq.~(\ref{eq:finiteT}) in a temperature regime where inelastic processes are favored by $k_{\mathrm{B}}T$ exceeding the characteristic energy scale for other degrees of freedom.  For example, at 300~K, the lowest optical phonon mode in Au ($\sim$~17~meV\cite{Kumar:2012}) should already be populated.

In this paper, we consider the noise properties of ensembles of atomic-scale metal point contacts from tunneling to the multichannel ($G \sim 6 G_{0}$) regime, at room temperature, when inelastic processes involving phonons should be considerably more important than in the cryogenic limit.   Of particular interest are the accuracy and utility of Eq.~(\ref{eq:finiteT}) under these conditions, over a broad range of applied bias, and the relative contributions of other noise mechanisms, such as conductance fluctuations\cite{Weissman:1988}.   With biases ranging from $1 < eV/k_{\mathrm{B}}T < 10$, we find noise consistent with Eq.~(\ref{eq:finiteT}), with clear relative suppression of the noise at conductance values corresponding to the quantized conductance peaks in the ensemble histograms.  At still lower bias, we cannot resolve the excess noise, while conductance peaks remain still clear.  The Fano factor in the high conductance, high bias regime is approximately a third of that in the $G < G_{0}$ tunneling regime.   At the conductances where noise is relatively suppressed, the bias scaling of the averaged noise is consistent with conductance fluctuations, and the magnitude is not unreasonable considering previous experiments\cite{Wu:2008}.

\begin{figure}[h]            
\includegraphics[width=3.4in]{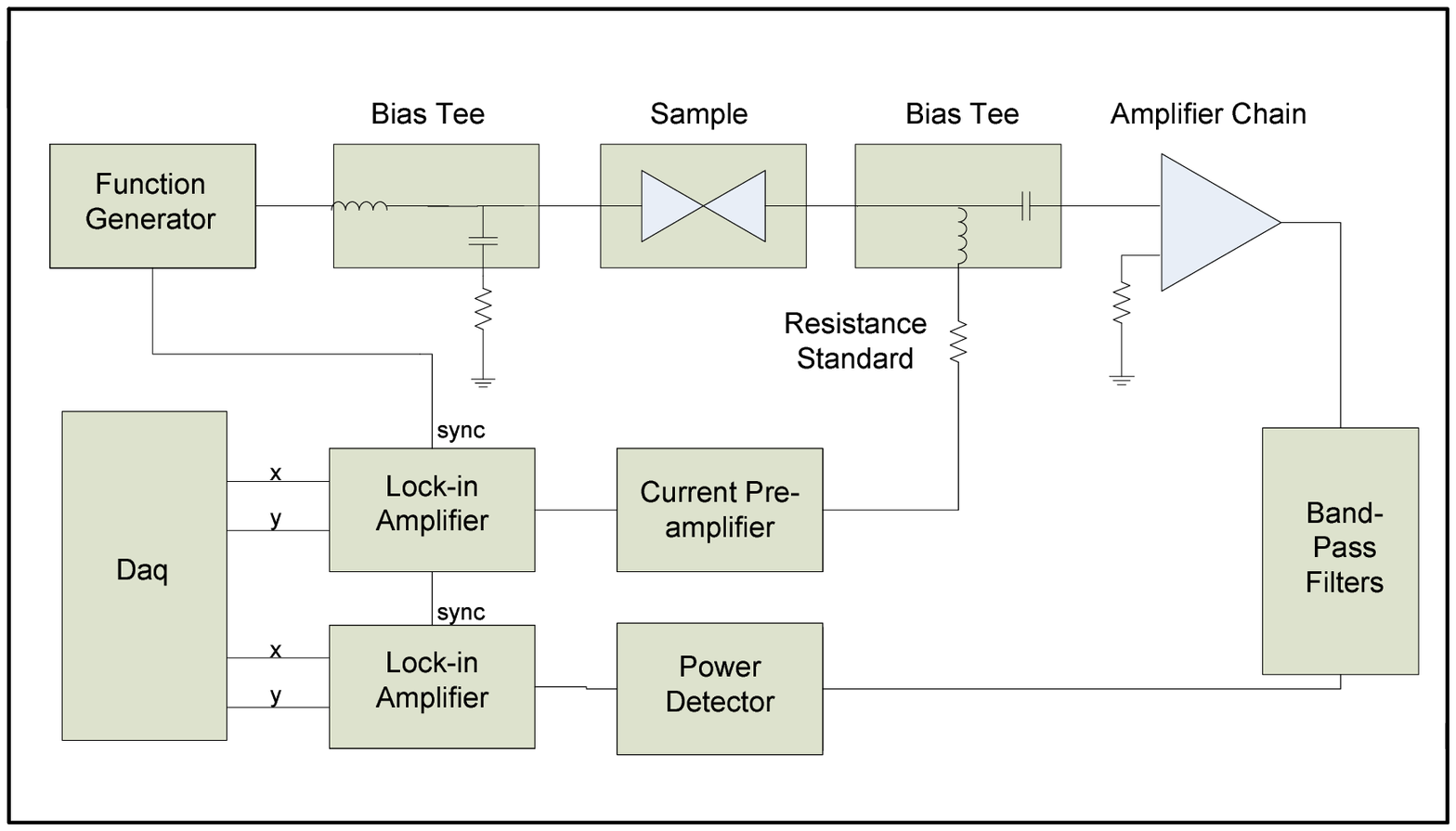} 
\caption{The schematic of the circuit used to measure DC conductance and RF noise power. All the RF components in this circuit have 50~$\Omega$ impedance except the sample.  }\label{fig:1} 
\end{figure} 

These experiments are performed using a scanning tunneling microscope (STM)-style break junction, as has become very popular in the study of molecular conduction\cite{Xu:2003,Venkataraman:2006}.   A junction is repeatedly made and broken in ambient conditions between a 50~nm-thick gold film evaporated on an oxidized silicon substrate, and a cut gold wire.  A computer-controlled piezo actuator is used to form and break the junction typically several times per second.  The noise measurement approach is similar that employed previously in a flexural mechanical break junction\cite{Wheeler:2010}.  The desire to examine the ensemble-averaged noise leads to the choice of the STM break junction method; the need for rapid measurement of the noise during the junction breaking process necessitates the use of a high bandwidth radio frequency (RF) technique.    Throughout the junction formation and breaking cycle, a ``DC'' bias square wave (between 0 V and a desired voltage level, with a frequency of approximately 10~kHz) is applied across the series combination of the gold junction and a current-limiting 2~K$\Omega$~ resistance standard.  This low frequency square wave serves as the (essentially) DC bias that drives current through the junction.  The  circuit, as shown in Fig. 1, employs bias-tees to separate the DC and RF signals coming from the junction.  A current preamplifier measures the current and is recorded electronically, giving a measure of the junction's conductance.   At the same time, a lock-in amplifier synchronized to the square wave detects the difference between the RF power with and without bias applied to the junction; this is the excess noise power.  The bandwidth of the noise measurement is roughly 250-580~MHz.  A detailed gain-bandwidth product measurement is employed.  Since both shot noise and Johnson-Nyquist noise are expected to be white over this bandwidth, deviations from white noise arise from the impedance properties of the measurement circuit as a whole.  The details of the noise analysis and background subtraction are described in supplemental material.

At every single bias, the STM style motion of the gold tip repeats hundreds times to generate a histogram of conductance, as well as a plot of the ensemble averaged excess noise power vs. conductance. An example is provided in Fig.2. 
\begin{figure}[h]            
\includegraphics[width=3.5in]{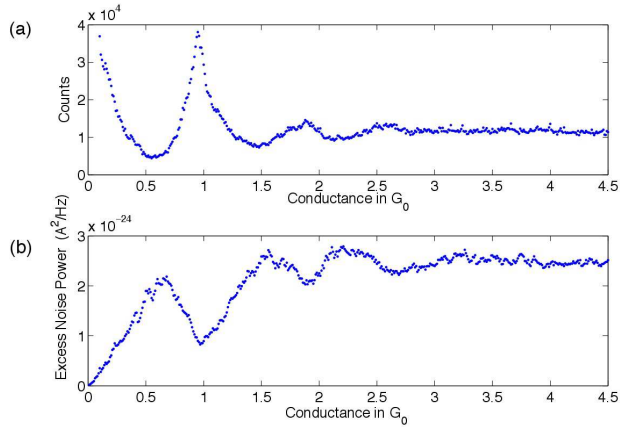} 
\caption{An example of conductance quantization (a) and noise suppression (b) at room temperature.  (a)  A conductance histogram acquired with 180 mV amplitude for the biasing square wave. (b) Averaged power spectral density vs. conductance.  The bin size in both plots is 0.01 $G_0$. }\label{fig:2} 
\end{figure} 
As has been seen in many previous experiments in atomic-scale metal contacts\cite{Agrait:2003}, peaks are observed in the conductance histograms, signifying preferred junction configurations with specific values of conductance.  Peaks are observed at 1 $G_{0}$, and near other integer multiples of $G_0$, consistent with past results on Au junctions at room temperature\cite{Yanson:2005}.   Cryogenic experiments involving shot noise\cite{vandenBrom:1999} and subgap structure in superconducting contacts\cite{Scheer:2001} have demonstrated that the 1 $G_{0}$ peak in Au junctions is dominated by configurations with a single highly transmitting channel ($\tau \rightarrow 1$).   In our structures usually the first three conductance peaks are readily resolvable in the histogram.  The related ensemble-averaged noise power measurement is also shown.  As is clear from the figure, the ensemble-averaged noise power is clearly suppressed near conductance values where the conductance histogram is peaked.

The transmission of RF signals always faces the problem of power reflection, which originates from impedance mismatch.   Conversely, reflection itself carries information about impedance.  In our measurement circuit, all the commercial RF electronic components are of 50$\Omega$ impedance.  We therefore expect significant impedance mismatch and reflections only between the STM-style gold junction and the transmission lines.  As an added complication compared to a fixed device configuration, the tip's repeatedly vertical motions introduce the extra complexity of a strongly time-varying DC conductance into the junction's RF properties.  In principle a measurement should be performed to properly characterize the impedance mismatch between the junction and the RF measurement circuitry at each conductance value.  Ideally, knowing the RF properties of the nanoscale junction and the accompanying electronics, including the gain-bandwidth product of the amplifier chain,  it should be possible to infer the actual current noise (A$^{2}$/Hz across the junction) from the measured RF power seen by the power meter.  However, in the STM breakjunction setup, in which the DC conductance of the junction changes by orders of magnitude on millisecond timescales, with our equipment it is not possible to measure all of the relevant RF parameters in real time.  As an approximation to this, we instead measure the reflection properties as a function of conductance \textit{averaged} over the ensemble of junction configurations.  This should at least indicate whether there are gross variations in the efficiency of the junction's RF coupling to the rest of the circuit.  

Following on the approach reported previously in measurements of the impedance properties of a vacuum photodiode\cite{Wheeler:2010}, we perform a reflectance measurement as shown in Fig. 3.  A commercial white noise source is used to provide wide-band white noise across the RF bandwidth of interest.   The amplitude of this noise is modulated by a subsequent RF switch that is turned on and off at the same (acoustic) frequency used for the square wave voltage bias we applied in excess noise measurement.   Part of the white noise will be reflected at the boundary between the gold junction and the transmission line to the bias tee. The reflected power goes through a directional coupler as well as the same amplifier chain used in the excess noise measurements, and is registered by the logarithmic power detector.  Simultaneously, the junction's cyclical STM-style motion is executed, with an applied DC bias across the junction to allow the simultaneous acquisition of a conductance histogram. This measurement gives a picture of the relation between ensemble averaged reflections and the conductance of the junction.  While the reflection measurement cannot provide all the information about the junction's RF properties, at least it provides a rough check that nothing dramatic happens in terms of the ensemble-averaged impedance mismatch over the conductance range.  The reflection averaged over the ensemble of junction configurations is around 22\%.  There is some systematic variation with conductance, but this is small, less than 2\% over 10~$G_0$.   

Since the impedance mismatch between the junction and the measurement circuit does not vary dramatically on average over the range of junction configurations, there should be a scale factor (approximately constant across the conductance range) between the measured power and the true current noise across the junction.  We attempt to find this factor by using the knowledge that the Fano factor in the tunneling regime $G << G_{0}$ approaches one in the limit that a single poorly transmitting channel dominates the conductance, as apparent from Eq.~(1).

\begin{figure}[h]            
\includegraphics[width=3.5in]{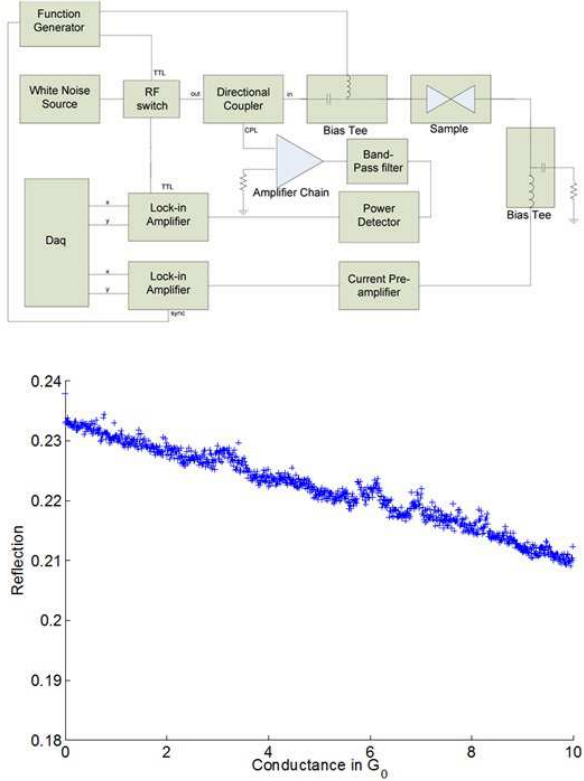} 
\caption{Top:  Schematic of reflection measurement setup.  Bottom: Measured, ensemble- and bandwidth-averaged reflection over the whole conductance range.}\label{fig:3} 
\end{figure} 

We acquire conductance histograms and ensemble-averaged noise using a series of square wave bias voltages from a few millivolts to several hundreds of millivolts.  An example of such ensemble-averaged noise data for eight different bias levels is shown in Fig.4.  All the data shown in this figure were taken continuously in one day to ensure an identical experimental environment (lab temperature, any stray RF background).  In our system no clear excess noise can be detected at bias voltages below about 25~mV, corresponding to $k_{\mathrm{B}}T$ at $T \approx 300$~K.  Noise suppression at conductances corresponding to the first three peaks in the conductance histograms are very clear, and the magnitude of the detected noise power is monotonously increasing with bias as expected.  Note that the suppressions are not complete.  This indicates that at least some of the junction configurations corresponding to peak conductance values result from a mixture of multiple, partially transmitting quantum channels.

\begin{figure}[h]            
\includegraphics[width=3.5in]{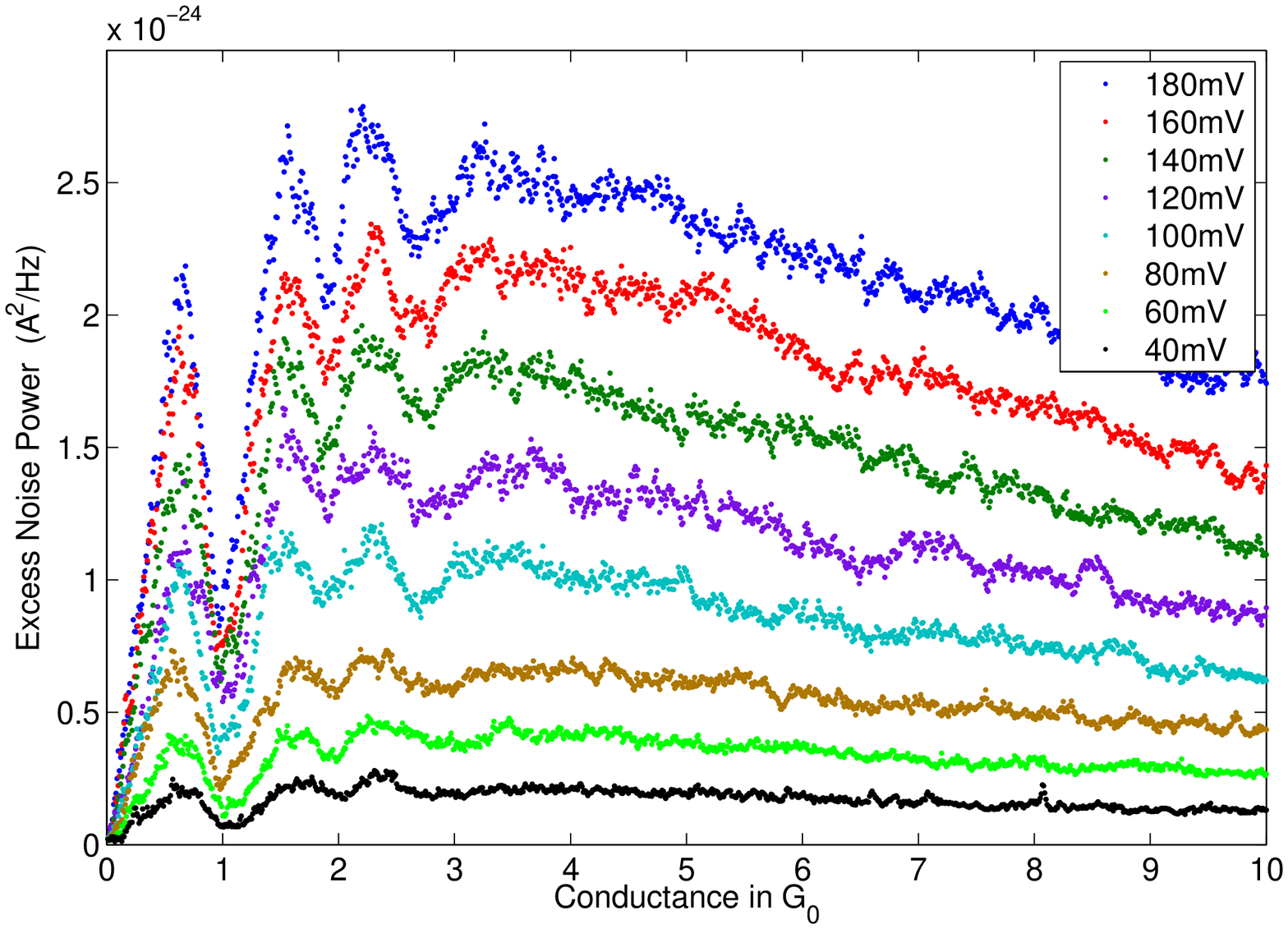} 
\caption{Excess noise spectral density vs conductance at different biases. Colors represent different bias voltages applied across the series combination of the junction and the resistance standard. }\label{fig:4} 
\end{figure}  

We are interested in the scaling of the noise with bias, as this reveals the Fano factor.  Working with the data sets in Fig. 4, we can specify a particular conductance value, and for each applied square wave bias voltage we can compute the actual voltage drop, $V$, across the junction.  Starting from Eq.~(\ref{eq:finiteT}), we can see that plotting the detected power spectral density as a function of $X=4k_{\mathrm{B}}TG[(eV/2k_{\mathrm{B}}T) \coth(eV/2k_{\mathrm{B}}T)-1]$ should produce a linear graph, with the slope giving the (zero temperature) Fano factor, $F$.  Here $G$ is the conductance of the junction and $T$ is assumed to be the ambient temperature, $\approx 300$~K.  In Fig. 5a, we plot data for three closely spaced values of conductance, all below $G_{0}$.  The figure shows that the noise is in fact linear when plotted as a function of $X$, and the Fano factor is decreasing as the conductance approaches 1~$G_{0}$.  These are consistent with the measured signal originating from shot noise, and the suppressions shown in Fig. 4a originating from the saturated channel mechanism of Eqs.~(\ref{eq:zeroT},~\ref{eq:finiteT}).  Qualitatively similar declines in Fano factor are also observed as the conductance approaches the values associated with the other two peaks in conductance histograms/suppressions in measured noise power.  

We now consider the situation at higher conductances, when the number of channels involved in transport is large compared to one.  As discussed above, we assume that the Fano factor in the low conductance regime is close to one, as expected if transport is dominated by a single poorly transmitting channel.   We perform a linear fit to the noise vs. $X$ data at $G \approx 0.1~G_{0}$, and assume that corresponds to a true Fano factor close to one (though a true single channel device would have $F \approx 0.9$ near that conductance).   In the particular data set shown in Fig.~5a, the lowest conductance data around $G = 0.45~G_{0}$ are still well described by a similar Fano factor, as shown by the blue line.  This suggests that, when looked over the full ensemble average, the conductance in this regime results from several poorly transmitting channels, rather than being dominated by a single channel with a transmittance of 0.45.  This indicates that a large contribution to the conductance histogram results from junctions with comparatively blunt tips.  Figure 5b shows, on the same plot as that low conductance data, the noise data vs. $X$ at around 4,5 and 6$G_0$.  Identically scaled linear fits show slopes of 0.39, 0.34, and 0.31, respectively.   These relative slopes are comparatively insensitive to the choice of the low conductance fits used to find the scaling factor, provided $G < \sim 0.5~G_{0}$.  At even higher conductance values not shown in the figure, the inferred Fano factors also do not vary by much, decreasing to around 0.25 near 10~$G_{0}$.   

These observations are roughly consistent with expectations for a crossover toward diffusive conduction.   In diffusive conductors that are small compared to the inelastic scattering length for both electron-electron ($\ell_{e-e}$) and electron-phonon ($\ell_{e-ph}$) interactions, theoretical calculations\cite{Beenakker:1992,Shimizu:1992,Nagaev:1992,deJong:1995} and experiments on different samples at helium temperature\cite{Henny:1999}  agree that $F \rightarrow 1/3$ in this regime.  In the limit that the constriction length $L$ exceeds $\ell_{e-e}$, the expected reduction factor is predicted\cite{Nagaev:1995} to be $\sqrt{3}/4 \approx 0.433$. Since both longitudinal and transverse dimensions in our junctions are comparable to the atomic scale, it seems likely that $L << \ell_{e-e},~\ell_{e-ph}$, and our data show consistency with the prediction of 1/3. This consistency with 
 a diffusive picture is a bit surprising, given that the length and transverse dimensions of the point contact should still be around 1 nm even when $G \approx 10~G_{0}$.  Experiments involving Pb junctions\cite{Riquelme:2005} have shown a similar transition to the diffusive regime even at relatively low conductances, though in that work this is ascribed to the importance of multiple orbitals per atom contributing to the conductance, leading to enhanced channel mixing.   While this multiorbital mechanism should not apply for Au junctions, the lack of clear peaks in the conductance histograms (and corresponding suppressions in the noise) above 3~$G_{0}$ are consistent with enhanced channel mixing in our structures relative to that seen in other Au experiments\cite{Ludoph:2000,Yanson:2005}.  This is further supported by the observation mentioned in the previous paragraph that conduction in the tunneling regime in this data set appears to involve multiple poorly transmitting channels.  It has been pointed out before\cite{Henny:1999} that the prediction of a suppression factor of about 1/3 is surprisingly robust.  Investigations with different materials and over a broader range of temperature and bias should shed light on the extent of universality in this suppression.

\begin{figure}[h]                                                                     
{
\includegraphics[width=0.5\textwidth]{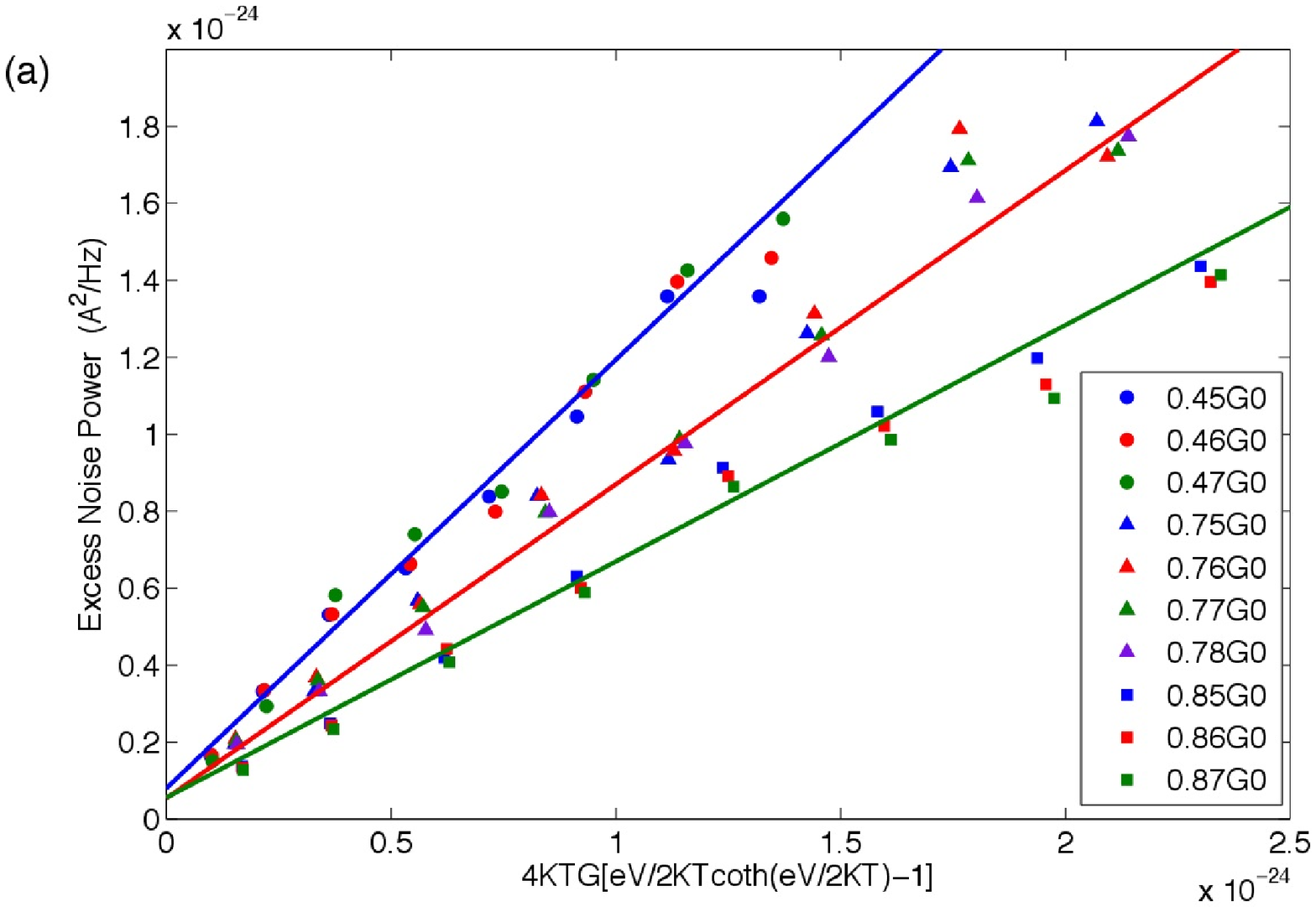}}
{
\includegraphics[width=0.5\textwidth]{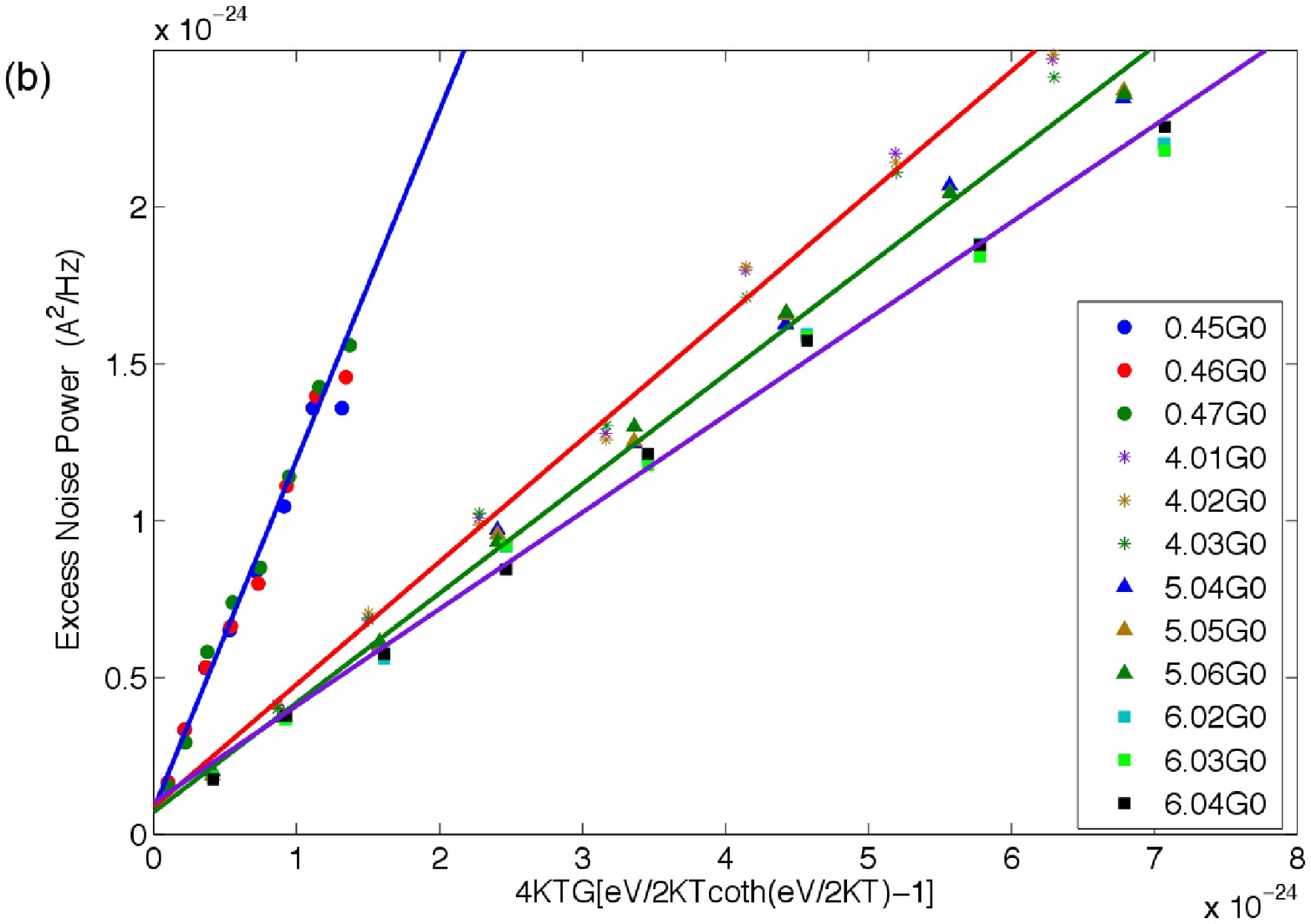}}
\caption{Noise as a function of scaled bias, for all eight data sets shown in Fig. 4. Top panel:  Data below 1 $G_0$. Data points at around 0.45~$G_0$, 0.78~$G_0$ and 0.85~$G_0$ are shown in the plot as well as linear fits to $X$ for those points.  Bottom panel:   Data around 4~$G_0$, 5~$G_0$ and 6~$G_0$, with inferred Fano factors (relative to that at $\sim 0.1~G_0$) of 0.39, 0.34 and 0.31 separately.}
\label{Fig.5}
\end{figure}

\begin{figure}[h]                                                                     
{
\includegraphics[width=0.5\textwidth]{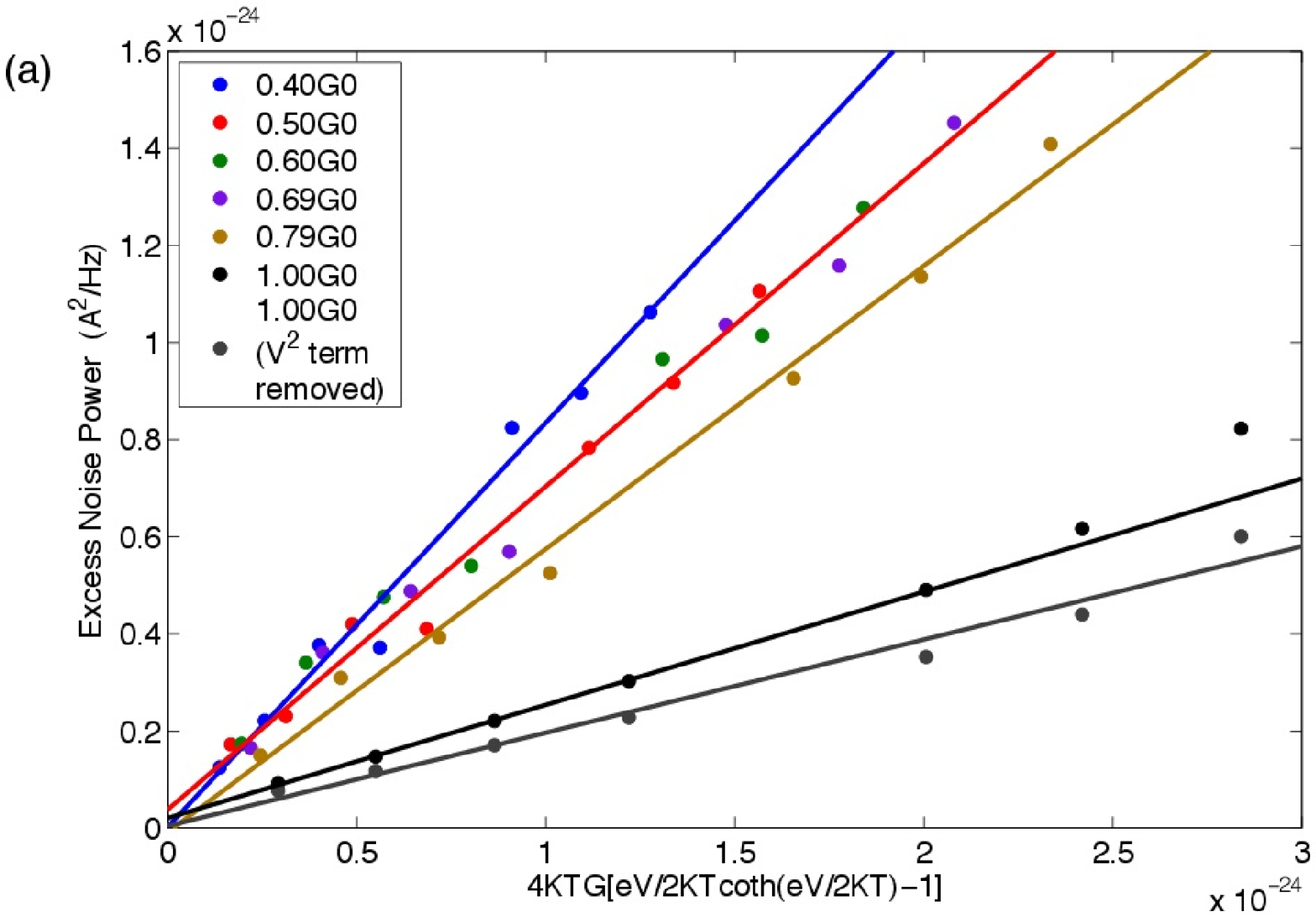}}
{
\includegraphics[width=0.5\textwidth]{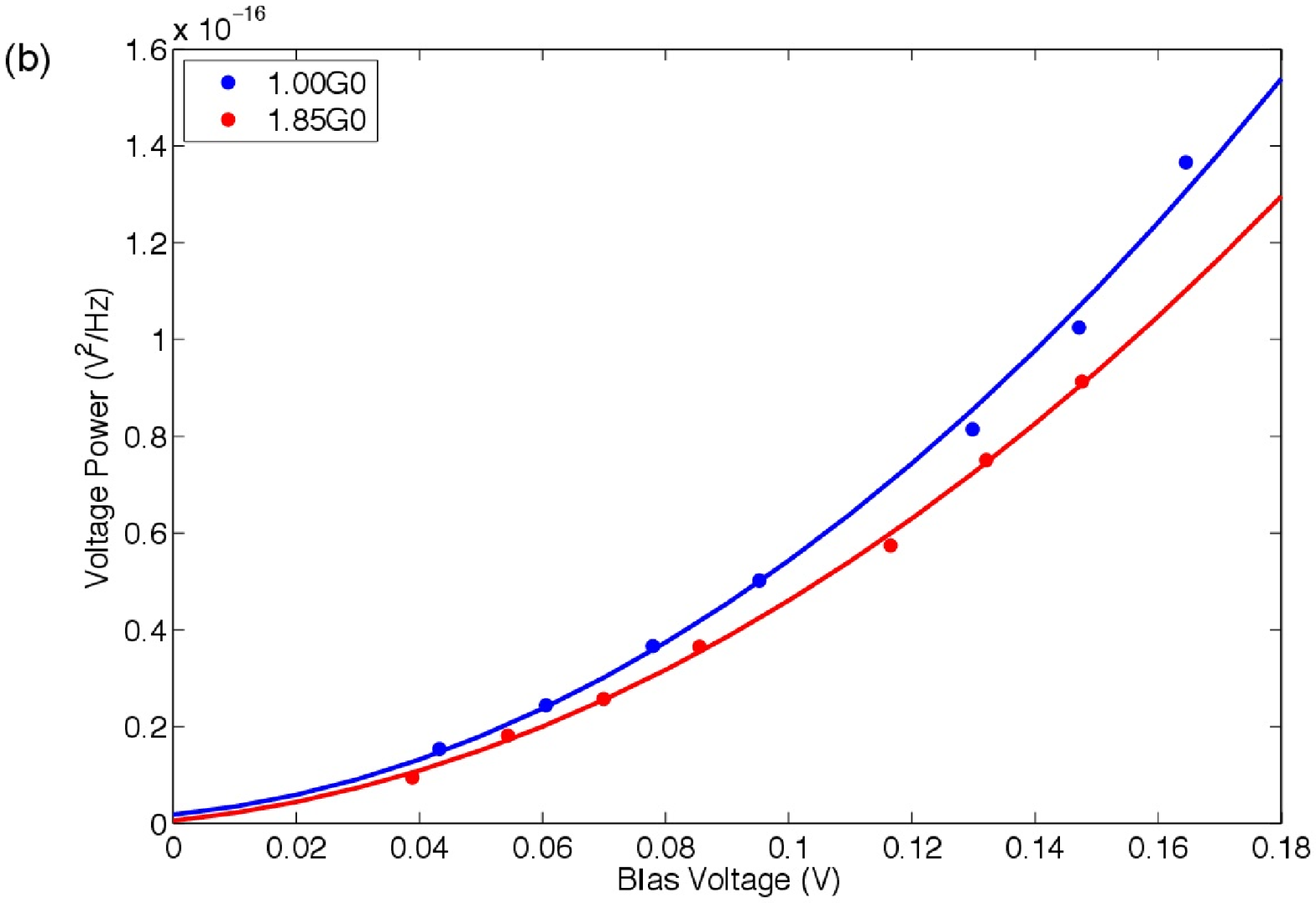}}
\caption{In the upper panel, off-suppression data and on-suppression noise data are compared. The former still has good linearity as a function of $X$.  However, for the latter a deviation from a linear dependence is observed when the conductance is around 1~$G_0$ and the shot noise contribution is relatively suppressed. The lowest (gray) points and linear fit correspond to the 1~$G_{0}$ data with a best-fit quadratic-in-voltage contribution removed.   Panel (b) is a purely quadratic fit of the excess noise vs. bias voltage at the conductances corresponding to the first two suppressions. The noise power here has been converted to mean squared voltage per frequency, for comparison with flicker noise.  Both of these fitting procedures give flicker noise contributions of the same order of magnitude.}
\label{Fig.6}
\end{figure}

We also consider the scaling of the excess noise at conductance values where the shot noise contribution is expected to be maximally suppressed.   We have taken data sets of the type shown in Fig.~4 on multiple occasions with different tips, film samples, and cleaning/annealing procedures.  To best study the bias scaling of the remaining excess noise near 1~$G_{0}$, we examine a data set when the suppression at 1~$G_{0}$ was particularly well defined.  The result is shown in Fig. 6a.   With the comparatively strong suppression of shot noise power in this data set, we find that the measured noise vs. $X$ is nonlinear.  This nonlinearity is only clearly seen at conductances where the noise is relatively suppressed.  A natural explanation for this is a contribution to the measured noise from conductance fluctuations, commonly termed ``flicker'' noise, which often has (in macroscale systems) a $1/f$ frequency dependence\cite{Wu:2008}.  Such flicker noise, originating with fluctua
 tions in the actual junction resistance, should scale quadratically with voltage across the junction at a given conductance, when recast as voltage fluctuations.  Figure 6b plots the excess noise spectral density vs bias voltage at the first two suppressions, with the noise power converted to mean squared voltage per unit frequency.  Quadratic fits of the form $S_{V}  = AV^{2} +BV+ C$ describe the data well, with a comparatively small linear term, $B$, as well as an even more tiny residual intercepts, $C$.  With this fitting procedure we find $A \approx 4 \times 10^{-15}$~Hz$^{-1}$.  We note that such a fitting procedure overestimates the size of $A$, since the hyperbolic cotangent term in Eq.~(\ref{eq:finiteT}) does contribute some nonlinearity in such a plot even when the only noise is finite temperature shot noise.  Instead if we fit a quadratic in voltage term in addition to the finite temperature shot noise expectation of Eq.~\ref{eq:finiteT}, we find $A \approx 1.4 \times 10^{-15}$~Hz$^{-1}$, smaller but of the same order of magnitude.   Using the phenomenological Hooge's law\cite{Weissman:1988}, 
\begin{equation}
S_R(f)/R^2=\frac{\alpha}{Nf^{\nu}}
\end{equation}
with power $\nu$ ranging between $1\sim2$, we compared our inferred magnitude of $A$ with values observed at lower frequencies and higher junction conductances in other gold point contacts by Wu \textit{et al.}\cite{Wu:2008}.  The result depends strongly on the assumed value of $\nu$, which is expected to fall between 1 (traditional $1/f$ noise) and 2 (expected for a single two-level fluctuator\cite{Wu:2008}).  We find a flicker noise amplitude smaller than the Wu \textit{et al.} values extrapolated to 1~$G_{0}$ with the fitted $\nu$ value in their case; however, if $\nu \approx 1.7$ over frequency from their sub-100~kHz measurements to our RF scale, our value for $A$ is compatible with their results.  We can reasonably conclude that much of the nonlinearity in the noise vs. bias at conductances when shot noise is suppressed results from flicker noise/conductance fluctuations.  The relatively small magnitude of this noise suggests a comparatively rapid decay of flicker noise with frequency in the RF range in atomic-scale junctions.  It would be interesting to consider the temperature variation of the flicker noise magnitude over this frequency range, which would help clarify whether thermal activation of defect motion is relevant here.

\begin{figure}[h]                                                                     
{
\includegraphics[width=0.5\textwidth]{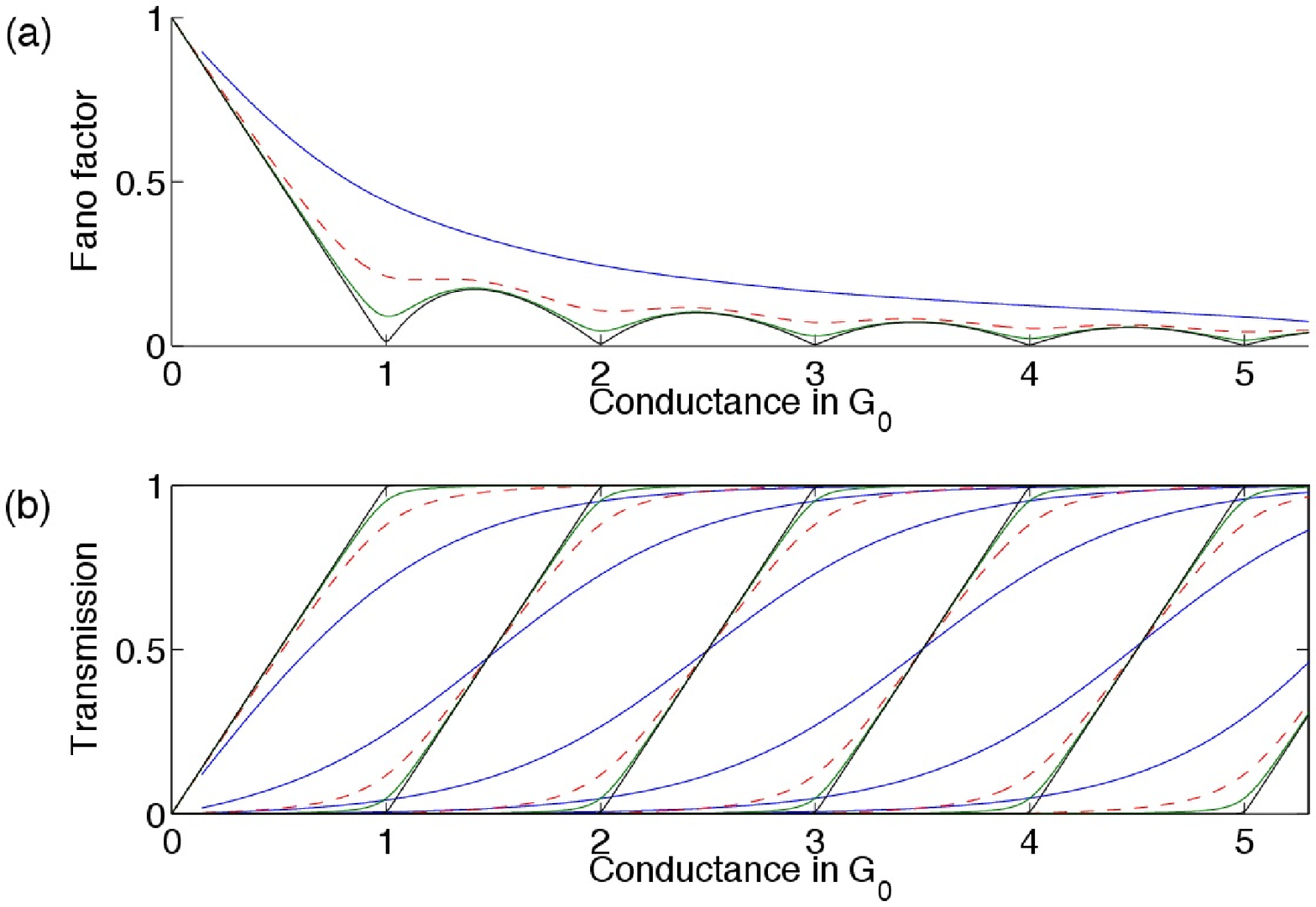}}
{
\includegraphics[width=0.5\textwidth]{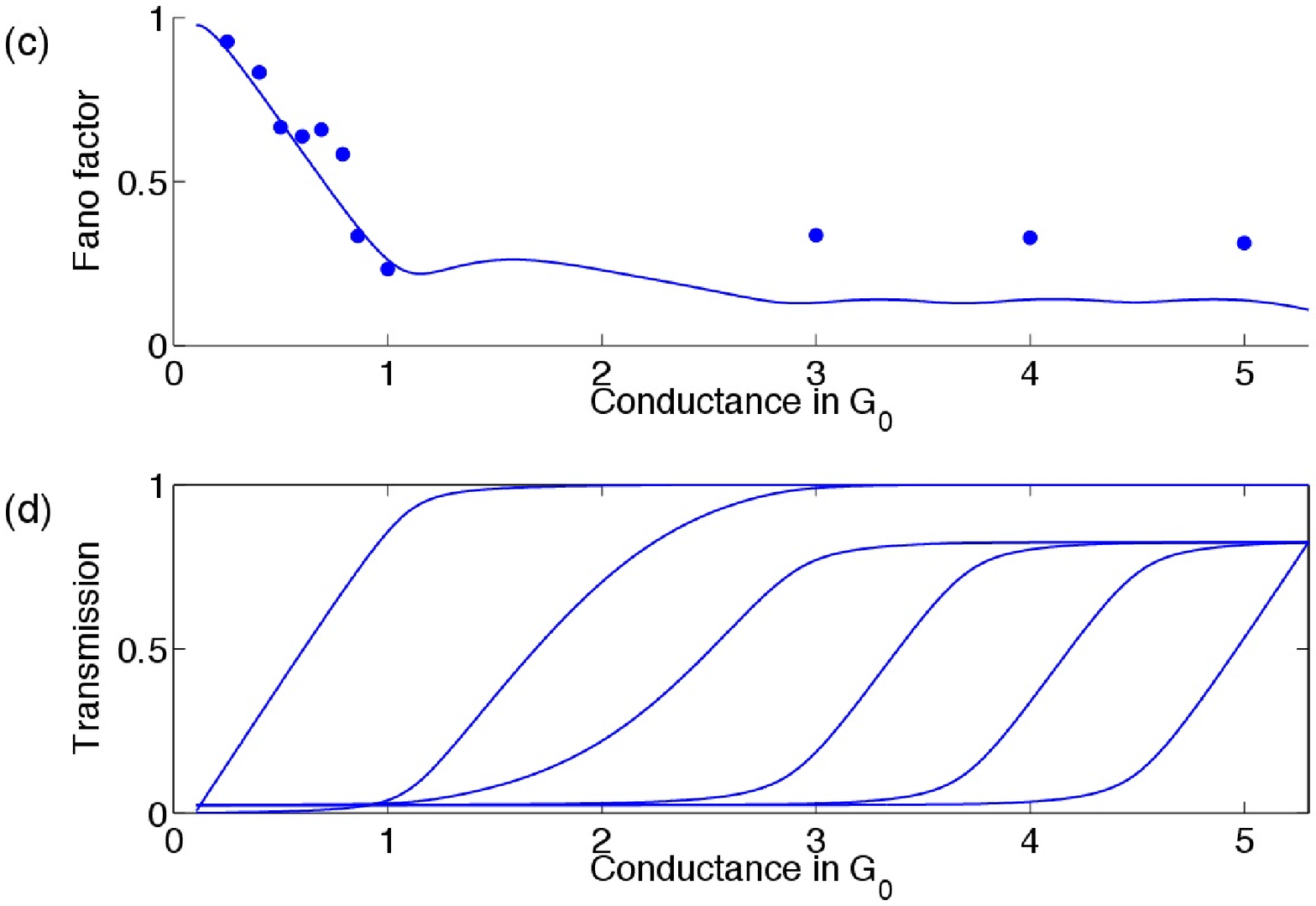}}
\caption{Models of channel mixing and Fano factors.  The top panel is the expected Fano factor as a function of conductance for the particular channel mixing shown in the second panel.   The third panel shows the measured Fano factor for a particular set of data, and a theoretical approximation to that data based on the channel openings shown in the bottom panel.  Colors and linestyles represent different evolutions of $\{\tau_i\}$ as a function of conductance. }
\label{Fig.7}
\end{figure}

It is worth considering what kind of mixing of quantum channels is required to produce the evolution of Fano factor with conductance that we infer from our data, within the noninteracting picture of Eq.~\ref{eq:zeroT}.   A simple ``toy model'' simulation of possible Fano factor evolution is provided, which depends on the choice of how the channels $\{\tau_i\}$ open as conductance is increased.  The result is shown in Fig.7 (a).  We assume for simplicity a set of uniformly separated hyperbolic tangent functions for the form of  $\{\tau_i\}$ as a function of $G$, which is shown on the bottom.  Each hyperbolic tangent is specified by a central conductance at which $\tau_{i} = 0.5$, and a width in $G$ over which the channel ``turns on''.   The resulting Fano factors have been plotted on the top.  Multiple colors/linestyles represent different choices of the central conductances and turn-on widths.  In top graph, the uppermost line (blue) shows no clear suppression of noise vs. $G
 $ due to the heavy overlaps between channels.  In contrast, the lower  most curve (black) is in the limit of well-separated channels that do not mix.  Channels' mixture tends to smear out suppressions and raise the Fano factor, which is the qualitative reason that in a larger-sized diffusive conductor only a roughly constant Fano factor results.   Figure 7b shows an example (nonunique), of a set of transmission channels that agree reasonably well with the measurements.  Additional constraints (such as those provided by subgap conductance in superconducting junctions\cite{Scheer:1998,Scheer:2001}) are necessary to better constrain the specific channels and their evolution. 

We note that we do not see clear signatures of bias-dependent changes in the Fano factor in these ensemble averaged measurements.  As the low temperature data in Kumar {\it et al.} shows\cite{Kumar:2012}, changes in the apparent Fano factor due to excitation of optical phonons can be of either sign in single-channel devices, with a crossover between enhancement and suppression of $F$ taking place at a particular transmittance, around 0.95.  With respect to the Au optical phonon at $\sim$~17~meV, all of the data in this paper are in the high bias regime above the voltage threshold.  Our measurements show that in Au junctions at room temperature, there are no clear inelastic thresholds above this energy that survive ensemble averaging over junction configurations.  Other analysis approaches that look at subensembles may be more revealing\cite{Makk:2012} and are in progress.

Here we have reported a calibrated measurement of shot noise and its bias dependence in STM-style gold junctions at room temperature.  We observe in general that the shot noise spectral density is proportional to $X \equiv 4k_{\mathrm{B}}TG[eV/2k_{\mathrm{B}}T \coth(eV/2k_{\mathrm{B}}T)-1]$, as expected for a nanoscale junction with a small number of channels.   The slope of such a plot is a means of extracting the zero-temperature Fano factor of the noise.  The Fano factor observed at higher conductances (several conductance quanta) is roughly 1/3 that seen in the tunneling regime, consistent with expectations as the diffusive limit is approached.  At suppressions where shot noise is comparatively small, nonlinearities in the noise vs. $X$ are observed.  It is likely that the origin of this nonlinearity is conductance fluctuation/flicker noise; this is supported by fits showing an approximate quadratic dependence of the noise on the bias across the junction, and the magnitude of this noise is roughly consistent with that extrapolated from other experiments\cite{Wu:2008}.   The evolution of the Fano factor with conductance can be described adequately within the noninteracting picture through reasonable choices of the evolution of channel numbers and transmittances with conductance.   These experiments highlight that the quantum nature of the electronic conduction remains detectable and important as electronic systems approach the atomic scale, even at room temperature.

The authors acknowledge the support of NSF award DMR-0855607, and useful conversations with J. van Ruitenbeek, C. Sch{\"o}nenberger, and L. Richardson.

\section*{Appendix:  Background Subtraction}

Scanning tunneling microscope-style break junctions provide us with a way to sample a wide range of conductance values quickly.   However, when using a lock-in measurement of the conductance, the repeated change of the conductance over orders of magnitude also causes rapid changes in the phase of the conductive (as opposed to displacement) contribution to the current, and therefore the noise signal.  As a practical matter it is not possible to adjust the phase of the lock-in measurement ``on the fly'' to separate out the conductive and displacement contributions to the current and the noise.  Instead, in our measurement the lock-in was set to measure the rms magnitude $R$ of the output of the power detector.   

Unfortunately, this introduces an effective background.  To see this, recall that $R=\sqrt{(X^2+Y^2)}$, where $X$, $Y$ are the two orthogonal components of the measured signal.  Even without any input signal from the power detector $X$ and $Y$ fluctuate about zero due to amplifier noise.  As a result, $R$ averages to a non-zero positive value even in the absence of a real noise signal, and this value is not negligible in our measurement.  During a real measurement, $R$ can be expressed as $R=\sqrt{(X+X_r)^2+(Y+Y_r)^2}$, where $X_r$ and $Y_r$ represent the random errors (amplifier noise, fluctuating about zero) on the two components. To remove this background on average, we have to perform one extra measurement at essentially zero bias, to characterize the background.  Practically we still use a very low bias voltage, usually around 3 millivolts, which is small enough that no true excess noise is detectable at room temperature, but we are still able to measure conductance.  In
  principle the output of the noise measurement lock-in during this measurement gives $R_0=\sqrt{X_r^2+Y_r^2}$.  After average $\langle R_0^2 \rangle =\langle X_r^2\rangle+\langle Y_r^2\rangle$, while $\langle R^2 \rangle= \langle X^2 \rangle+ \langle X_r^2 \rangle + \langle Y^2 \rangle + \langle Y_r^2 \rangle$,  because the $\langle X_r \rangle$ and $\langle Y_r \rangle$ terms average to zero.  We can then consider $\langle R^2 \rangle - \langle R_0^2 \rangle$ as the averaged mean square signal from the power detector, which is the desired quantity.


\begin{thebibliography}{37}%
\makeatletter
\providecommand \@ifxundefined [1]{%
 \@ifx{#1\undefined}
}%
\providecommand \@ifnum [1]{%
 \ifnum #1\expandafter \@firstoftwo
 \else \expandafter \@secondoftwo
 \fi
}%
\providecommand \@ifx [1]{%
 \ifx #1\expandafter \@firstoftwo
 \else \expandafter \@secondoftwo
 \fi
}%
\providecommand \natexlab [1]{#1}%
\providecommand \enquote  [1]{``#1''}%
\providecommand \bibnamefont  [1]{#1}%
\providecommand \bibfnamefont [1]{#1}%
\providecommand \citenamefont [1]{#1}%
\providecommand \href@noop [0]{\@secondoftwo}%
\providecommand \href [0]{\begingroup \@sanitize@url \@href}%
\providecommand \@href[1]{\@@startlink{#1}\@@href}%
\providecommand \@@href[1]{\endgroup#1\@@endlink}%
\providecommand \@sanitize@url [0]{\catcode `\\12\catcode `\$12\catcode
  `\&12\catcode `\#12\catcode `\^12\catcode `\_12\catcode `\%12\relax}%
\providecommand \@@startlink[1]{}%
\providecommand \@@endlink[0]{}%
\providecommand \url  [0]{\begingroup\@sanitize@url \@url }%
\providecommand \@url [1]{\endgroup\@href {#1}{\urlprefix }}%
\providecommand \urlprefix  [0]{URL }%
\providecommand \Eprint [0]{\href }%
\providecommand \doibase [0]{http://dx.doi.org/}%
\providecommand \selectlanguage [0]{\@gobble}%
\providecommand \bibinfo  [0]{\@secondoftwo}%
\providecommand \bibfield  [0]{\@secondoftwo}%
\providecommand \translation [1]{[#1]}%
\providecommand \BibitemOpen [0]{}%
\providecommand \bibitemStop [0]{}%
\providecommand \bibitemNoStop [0]{.\EOS\space}%
\providecommand \EOS [0]{\spacefactor3000\relax}%
\providecommand \BibitemShut  [1]{\csname bibitem#1\endcsname}%
\let\auto@bib@innerbib\@empty
\bibitem [{\citenamefont {Schottky}(1918)}]{Schottky:1918}%
  \BibitemOpen
  \bibfield  {author} {\bibinfo {author} {\bibfnamefont {W.}~\bibnamefont
  {Schottky}},\ }\href@noop {} {\bibfield  {journal} {\bibinfo  {journal} {Ann.
  der Physik}\ }\textbf {\bibinfo {volume} {57}},\ \bibinfo {pages} {541}
  (\bibinfo {year} {1918})}\BibitemShut {NoStop}%
\bibitem [{\citenamefont {Johnson}(1928)}]{Johnson:1928}%
  \BibitemOpen
  \bibfield  {author} {\bibinfo {author} {\bibfnamefont {J.~B.}\ \bibnamefont
  {Johnson}},\ }\href {\doibase 10.1103/PhysRev.32.97} {\bibfield  {journal}
  {\bibinfo  {journal} {Phys. Rev.}\ }\textbf {\bibinfo {volume} {32}},\
  \bibinfo {pages} {97} (\bibinfo {year} {1928})}\BibitemShut {NoStop}%
\bibitem [{\citenamefont {Nyquist}(1928)}]{Nyquist:1928}%
  \BibitemOpen
  \bibfield  {author} {\bibinfo {author} {\bibfnamefont {H.}~\bibnamefont
  {Nyquist}},\ }\href {\doibase 10.1103/PhysRev.32.110} {\bibfield  {journal}
  {\bibinfo  {journal} {Phys. Rev.}\ }\textbf {\bibinfo {volume} {32}},\
  \bibinfo {pages} {110} (\bibinfo {year} {1928})}\BibitemShut {NoStop}%
\bibitem [{\citenamefont {Blanter}\ and\ \citenamefont
  {B{\"u}ttiker}(2000)}]{Blanter:2000}%
  \BibitemOpen
  \bibfield  {author} {\bibinfo {author} {\bibfnamefont {Y.}~\bibnamefont
  {Blanter}}\ and\ \bibinfo {author} {\bibfnamefont {M.}~\bibnamefont
  {B{\"u}ttiker}},\ }\href {\doibase 10.1016/S0370-1573(99)00123-4} {\bibfield
  {journal} {\bibinfo  {journal} {Physics Reports}\ }\textbf {\bibinfo {volume}
  {336}},\ \bibinfo {pages} {1 } (\bibinfo {year} {2000})}\BibitemShut
  {NoStop}%
\bibitem [{\citenamefont {Reznikov}\ \emph {et~al.}(1995)\citenamefont
  {Reznikov}, \citenamefont {Heiblum}, \citenamefont {Shtrikman},\ and\
  \citenamefont {Mahalu}}]{Reznikov:1995}%
  \BibitemOpen
  \bibfield  {author} {\bibinfo {author} {\bibfnamefont {M.}~\bibnamefont
  {Reznikov}}, \bibinfo {author} {\bibfnamefont {M.}~\bibnamefont {Heiblum}},
  \bibinfo {author} {\bibfnamefont {H.}~\bibnamefont {Shtrikman}}, \ and\
  \bibinfo {author} {\bibfnamefont {D.}~\bibnamefont {Mahalu}},\ }\href
  {\doibase 10.1103/PhysRevLett.75.3340} {\bibfield  {journal} {\bibinfo
  {journal} {Phys. Rev. Lett.}\ }\textbf {\bibinfo {volume} {75}},\ \bibinfo
  {pages} {3340} (\bibinfo {year} {1995})}\BibitemShut {NoStop}%
\bibitem [{\citenamefont {Kumar}\ \emph {et~al.}(1996)\citenamefont {Kumar},
  \citenamefont {Saminadayar}, \citenamefont {Glattli}, \citenamefont {Jin},\
  and\ \citenamefont {Etienne}}]{Kumar:1996}%
  \BibitemOpen
  \bibfield  {author} {\bibinfo {author} {\bibfnamefont {A.}~\bibnamefont
  {Kumar}}, \bibinfo {author} {\bibfnamefont {L.}~\bibnamefont {Saminadayar}},
  \bibinfo {author} {\bibfnamefont {D.~C.}\ \bibnamefont {Glattli}}, \bibinfo
  {author} {\bibfnamefont {Y.}~\bibnamefont {Jin}}, \ and\ \bibinfo {author}
  {\bibfnamefont {B.}~\bibnamefont {Etienne}},\ }\href {\doibase
  10.1103/PhysRevLett.76.2778} {\bibfield  {journal} {\bibinfo  {journal}
  {Phys. Rev. Lett.}\ }\textbf {\bibinfo {volume} {76}},\ \bibinfo {pages}
  {2778} (\bibinfo {year} {1996})}\BibitemShut {NoStop}%
\bibitem [{\citenamefont {Liefrink}\ \emph {et~al.}(1994)\citenamefont
  {Liefrink}, \citenamefont {Dijkhuis}, \citenamefont {de~Jong}, \citenamefont
  {Molenkamp},\ and\ \citenamefont {van Houten}}]{Liefrink:1994}%
  \BibitemOpen
  \bibfield  {author} {\bibinfo {author} {\bibfnamefont {F.}~\bibnamefont
  {Liefrink}}, \bibinfo {author} {\bibfnamefont {J.~I.}\ \bibnamefont
  {Dijkhuis}}, \bibinfo {author} {\bibfnamefont {M.~J.~M.}\ \bibnamefont
  {de~Jong}}, \bibinfo {author} {\bibfnamefont {L.~W.}\ \bibnamefont
  {Molenkamp}}, \ and\ \bibinfo {author} {\bibfnamefont {H.}~\bibnamefont {van
  Houten}},\ }\href {\doibase 10.1103/PhysRevB.49.14066} {\bibfield  {journal}
  {\bibinfo  {journal} {Phys. Rev. B}\ }\textbf {\bibinfo {volume} {49}},\
  \bibinfo {pages} {14066} (\bibinfo {year} {1994})}\BibitemShut {NoStop}%
\bibitem [{\citenamefont {Henny}\ \emph {et~al.}(1999)\citenamefont {Henny},
  \citenamefont {Oberholzer}, \citenamefont {Strunk},\ and\ \citenamefont
  {Sch\"onenberger}}]{Henny:1999}%
  \BibitemOpen
  \bibfield  {author} {\bibinfo {author} {\bibfnamefont {M.}~\bibnamefont
  {Henny}}, \bibinfo {author} {\bibfnamefont {S.}~\bibnamefont {Oberholzer}},
  \bibinfo {author} {\bibfnamefont {C.}~\bibnamefont {Strunk}}, \ and\ \bibinfo
  {author} {\bibfnamefont {C.}~\bibnamefont {Sch\"onenberger}},\ }\href
  {\doibase 10.1103/PhysRevB.59.2871} {\bibfield  {journal} {\bibinfo
  {journal} {Phys. Rev. B}\ }\textbf {\bibinfo {volume} {59}},\ \bibinfo
  {pages} {2871} (\bibinfo {year} {1999})}\BibitemShut {NoStop}%
\bibitem [{\citenamefont {van~den Brom}\ and\ \citenamefont {van
  Ruitenbeek}(1999)}]{vandenBrom:1999}%
  \BibitemOpen
  \bibfield  {author} {\bibinfo {author} {\bibfnamefont {H.~E.}\ \bibnamefont
  {van~den Brom}}\ and\ \bibinfo {author} {\bibfnamefont {J.~M.}\ \bibnamefont
  {van Ruitenbeek}},\ }\href {\doibase 10.1103/PhysRevLett.82.1526} {\bibfield
  {journal} {\bibinfo  {journal} {Phys. Rev. Lett.}\ }\textbf {\bibinfo
  {volume} {82}},\ \bibinfo {pages} {1526} (\bibinfo {year}
  {1999})}\BibitemShut {NoStop}%
\bibitem [{\citenamefont {Djukic}\ and\ \citenamefont {van
  Ruitenbeek}(2006)}]{Djukic:2006}%
  \BibitemOpen
  \bibfield  {author} {\bibinfo {author} {\bibfnamefont {D.}~\bibnamefont
  {Djukic}}\ and\ \bibinfo {author} {\bibfnamefont {J.~M.}\ \bibnamefont {van
  Ruitenbeek}},\ }\href {\doibase 10.1021/nl060116e} {\bibfield  {journal}
  {\bibinfo  {journal} {Nano Letters}\ }\textbf {\bibinfo {volume} {6}},\
  \bibinfo {pages} {789} (\bibinfo {year} {2006})}. \BibitemShut {NoStop}%
\bibitem [{\citenamefont {Saminadayar}\ \emph {et~al.}(1997)\citenamefont
  {Saminadayar}, \citenamefont {Glattli}, \citenamefont {Jin},\ and\
  \citenamefont {Etienne}}]{Saminadayar:1997}%
  \BibitemOpen
  \bibfield  {author} {\bibinfo {author} {\bibfnamefont {L.}~\bibnamefont
  {Saminadayar}}, \bibinfo {author} {\bibfnamefont {D.~C.}\ \bibnamefont
  {Glattli}}, \bibinfo {author} {\bibfnamefont {Y.}~\bibnamefont {Jin}}, \ and\
  \bibinfo {author} {\bibfnamefont {B.}~\bibnamefont {Etienne}},\ }\href
  {\doibase 10.1103/PhysRevLett.79.2526} {\bibfield  {journal} {\bibinfo
  {journal} {Phys. Rev. Lett.}\ }\textbf {\bibinfo {volume} {79}},\ \bibinfo
  {pages} {2526} (\bibinfo {year} {1997})}\BibitemShut {NoStop}%
\bibitem [{\citenamefont {de~Picciotto}\ \emph {et~al.}(1997)\citenamefont
  {de~Picciotto}, \citenamefont {Reznikov}, \citenamefont {Heiblum},
  \citenamefont {Umansky}, \citenamefont {Bunin},\ and\ \citenamefont
  {Mahalu}}]{dePicciotto:1997}%
  \BibitemOpen
  \bibfield  {author} {\bibinfo {author} {\bibfnamefont {R.}~\bibnamefont
  {de~Picciotto}}, \bibinfo {author} {\bibfnamefont {M.}~\bibnamefont
  {Reznikov}}, \bibinfo {author} {\bibfnamefont {M.}~\bibnamefont {Heiblum}},
  \bibinfo {author} {\bibfnamefont {V.}~\bibnamefont {Umansky}}, \bibinfo
  {author} {\bibfnamefont {G.}~\bibnamefont {Bunin}}, \ and\ \bibinfo {author}
  {\bibfnamefont {D.}~\bibnamefont {Mahalu}},\ }\href@noop {} {\bibfield
  {journal} {\bibinfo  {journal} {Nature}\ }\textbf {\bibinfo {volume} {389}},\
  \bibinfo {pages} {162} (\bibinfo {year} {1997})}\BibitemShut {NoStop}%
\bibitem [{\citenamefont {Wheeler}\ \emph {et~al.}(2010)\citenamefont
  {Wheeler}, \citenamefont {Russom}, \citenamefont {Evans}, \citenamefont
  {King},\ and\ \citenamefont {Natelson}}]{Wheeler:2010}%
  \BibitemOpen
  \bibfield  {author} {\bibinfo {author} {\bibfnamefont {P.~J.}\ \bibnamefont
  {Wheeler}}, \bibinfo {author} {\bibfnamefont {J.~N.}\ \bibnamefont {Russom}},
  \bibinfo {author} {\bibfnamefont {K.}~\bibnamefont {Evans}}, \bibinfo
  {author} {\bibfnamefont {N.~S.}\ \bibnamefont {King}}, \ and\ \bibinfo
  {author} {\bibfnamefont {D.}~\bibnamefont {Natelson}},\ } {\bibfield  {journal} {\bibinfo  {journal} {Nano Letters}\
  }}\textbf {\bibinfo {volume} {10}},\ \bibinfo {pages} {1287} (\bibinfo {year}
  {2010}). \BibitemShut {NoStop}%
\bibitem [{\citenamefont {Beenakker}\ and\ \citenamefont
  {B\"uttiker}(1992)}]{Beenakker:1992}%
  \BibitemOpen
  \bibfield  {author} {\bibinfo {author} {\bibfnamefont {C.~W.~J.}\
  \bibnamefont {Beenakker}}\ and\ \bibinfo {author} {\bibfnamefont
  {M.}~\bibnamefont {B\"uttiker}},\ }\href {\doibase 10.1103/PhysRevB.46.1889}
  {\bibfield  {journal} {\bibinfo  {journal} {Phys. Rev. B}\ }\textbf {\bibinfo
  {volume} {46}},\ \bibinfo {pages} {1889} (\bibinfo {year}
  {1992})}\BibitemShut {NoStop}%
\bibitem [{\citenamefont {Shimizu}\ and\ \citenamefont
  {Ueda}(1992)}]{Shimizu:1992}%
  \BibitemOpen
  \bibfield  {author} {\bibinfo {author} {\bibfnamefont {A.}~\bibnamefont
  {Shimizu}}\ and\ \bibinfo {author} {\bibfnamefont {M.}~\bibnamefont {Ueda}},\
  }\href {\doibase 10.1103/PhysRevLett.69.1403} {\bibfield  {journal} {\bibinfo
   {journal} {Phys. Rev. Lett.}\ }\textbf {\bibinfo {volume} {69}},\ \bibinfo
  {pages} {1403} (\bibinfo {year} {1992})}\BibitemShut {NoStop}%
\bibitem [{\citenamefont {Nagaev}(1992)}]{Nagaev:1992}%
  \BibitemOpen
  \bibfield  {author} {\bibinfo {author} {\bibfnamefont {K.}~\bibnamefont
  {Nagaev}},\ }\href {\doibase 10.1016/0375-9601(92)90814-3} {\bibfield
  {journal} {\bibinfo  {journal} {Physics Letters A}\ }\textbf {\bibinfo
  {volume} {169}},\ \bibinfo {pages} {103 } (\bibinfo {year}
  {1992})}\BibitemShut {NoStop}%
\bibitem [{\citenamefont {de~Jong}\ and\ \citenamefont
  {Beenakker}(1995)}]{deJong:1995}%
  \BibitemOpen
  \bibfield  {author} {\bibinfo {author} {\bibfnamefont {M.~J.~M.}\
  \bibnamefont {de~Jong}}\ and\ \bibinfo {author} {\bibfnamefont {C.~W.~J.}\
  \bibnamefont {Beenakker}},\ }\href {\doibase 10.1103/PhysRevB.51.16867}
  {\bibfield  {journal} {\bibinfo  {journal} {Phys. Rev. B}\ }\textbf {\bibinfo
  {volume} {51}},\ \bibinfo {pages} {16867} (\bibinfo {year}
  {1995})}\BibitemShut {NoStop}%
\bibitem [{\citenamefont {Steinbach}\ \emph {et~al.}(1996)\citenamefont
  {Steinbach}, \citenamefont {Martinis},\ and\ \citenamefont
  {Devoret}}]{Steinbach:1996}%
  \BibitemOpen
  \bibfield  {author} {\bibinfo {author} {\bibfnamefont {A.~H.}\ \bibnamefont
  {Steinbach}}, \bibinfo {author} {\bibfnamefont {J.~M.}\ \bibnamefont
  {Martinis}}, \ and\ \bibinfo {author} {\bibfnamefont {M.~H.}\ \bibnamefont
  {Devoret}},\ }\href {\doibase 10.1103/PhysRevLett.76.3806} {\bibfield
  {journal} {\bibinfo  {journal} {Phys. Rev. Lett.}\ }\textbf {\bibinfo
  {volume} {76}},\ \bibinfo {pages} {3806} (\bibinfo {year}
  {1996})}\BibitemShut {NoStop}%
\bibitem [{\citenamefont {Schoelkopf}\ \emph {et~al.}(1997)\citenamefont
  {Schoelkopf}, \citenamefont {Burke}, \citenamefont {Kozhevnikov},
  \citenamefont {Prober},\ and\ \citenamefont {Rooks}}]{Schoelkopf:1997}%
  \BibitemOpen
  \bibfield  {author} {\bibinfo {author} {\bibfnamefont {R.~J.}\ \bibnamefont
  {Schoelkopf}}, \bibinfo {author} {\bibfnamefont {P.~J.}\ \bibnamefont
  {Burke}}, \bibinfo {author} {\bibfnamefont {A.~A.}\ \bibnamefont
  {Kozhevnikov}}, \bibinfo {author} {\bibfnamefont {D.~E.}\ \bibnamefont
  {Prober}}, \ and\ \bibinfo {author} {\bibfnamefont {M.~J.}\ \bibnamefont
  {Rooks}},\ }\href {\doibase 10.1103/PhysRevLett.78.3370} {\bibfield
  {journal} {\bibinfo  {journal} {Phys. Rev. Lett.}\ }\textbf {\bibinfo
  {volume} {78}},\ \bibinfo {pages} {3370} (\bibinfo {year}
  {1997})}\BibitemShut {NoStop}%
\bibitem [{\citenamefont {Mitra}\ \emph {et~al.}(2004)\citenamefont {Mitra},
  \citenamefont {Aleiner},\ and\ \citenamefont {Millis}}]{Mitra:2004}%
  \BibitemOpen
  \bibfield  {author} {\bibinfo {author} {\bibfnamefont {A.}~\bibnamefont
  {Mitra}}, \bibinfo {author} {\bibfnamefont {I.}~\bibnamefont {Aleiner}}, \
  and\ \bibinfo {author} {\bibfnamefont {A.~J.}\ \bibnamefont {Millis}},\
  }\href {\doibase 10.1103/PhysRevB.69.245302} {\bibfield  {journal} {\bibinfo
  {journal} {Phys. Rev. B}\ }\textbf {\bibinfo {volume} {69}},\ \bibinfo
  {pages} {245302} (\bibinfo {year} {2004})}\BibitemShut {NoStop}%
\bibitem [{\citenamefont {Koch}\ and\ \citenamefont {von
  Oppen}(2005)}]{Koch:2005}%
  \BibitemOpen
  \bibfield  {author} {\bibinfo {author} {\bibfnamefont {J.}~\bibnamefont
  {Koch}}\ and\ \bibinfo {author} {\bibfnamefont {F.}~\bibnamefont {von
  Oppen}},\ }\href {\doibase 10.1103/PhysRevLett.94.206804} {\bibfield
  {journal} {\bibinfo  {journal} {Phys. Rev. Lett.}\ }\textbf {\bibinfo
  {volume} {94}},\ \bibinfo {pages} {206804} (\bibinfo {year}
  {2005})}\BibitemShut {NoStop}%
\bibitem [{\citenamefont {Koch}\ \emph {et~al.}(2006)\citenamefont {Koch},
  \citenamefont {von Oppen},\ and\ \citenamefont {Andreev}}]{Koch:2006}%
  \BibitemOpen
  \bibfield  {author} {\bibinfo {author} {\bibfnamefont {J.}~\bibnamefont
  {Koch}}, \bibinfo {author} {\bibfnamefont {F.}~\bibnamefont {von Oppen}}, \
  and\ \bibinfo {author} {\bibfnamefont {A.~V.}\ \bibnamefont {Andreev}},\
  }\href {\doibase 10.1103/PhysRevB.74.205438} {\bibfield  {journal} {\bibinfo
  {journal} {Phys. Rev. B}\ }\textbf {\bibinfo {volume} {74}},\ \bibinfo
  {pages} {205438} (\bibinfo {year} {2006})}\BibitemShut {NoStop}%
\bibitem [{\citenamefont {Leturcq}\ \emph {et~al.}(2009)\citenamefont
  {Leturcq}, \citenamefont {Stampfer}, \citenamefont {Inderbitzin},
  \citenamefont {Durrer}, \citenamefont {Hierold}, \citenamefont {Mariani},
  \citenamefont {Schultz}, \citenamefont {von Oppen},\ and\ \citenamefont
  {Ensslin}}]{Leturcq:2009}%
  \BibitemOpen
  \bibfield  {author} {\bibinfo {author} {\bibfnamefont {R.}~\bibnamefont
  {Leturcq}}, \bibinfo {author} {\bibfnamefont {C.}~\bibnamefont {Stampfer}},
  \bibinfo {author} {\bibfnamefont {K.}~\bibnamefont {Inderbitzin}}, \bibinfo
  {author} {\bibfnamefont {L.}~\bibnamefont {Durrer}}, \bibinfo {author}
  {\bibfnamefont {C.}~\bibnamefont {Hierold}}, \bibinfo {author} {\bibfnamefont
  {E.}~\bibnamefont {Mariani}}, \bibinfo {author} {\bibfnamefont {M.~G.}\
  \bibnamefont {Schultz}}, \bibinfo {author} {\bibfnamefont {F.}~\bibnamefont
  {von Oppen}}, \ and\ \bibinfo {author} {\bibfnamefont {K.}~\bibnamefont
  {Ensslin}},\ }\href {\doibase 10.1038/nphys1234} {\bibfield  {journal}
  {\bibinfo  {journal} {Nature Phys.}\ }\textbf {\bibinfo {volume} {5}},\
  \bibinfo {pages} {327} (\bibinfo {year} {2009})}\BibitemShut {NoStop}%
\bibitem [{\citenamefont {Fay}\ \emph {et~al.}(2011)\citenamefont {Fay},
  \citenamefont {Danneau}, \citenamefont {Viljas}, \citenamefont {Wu},
  \citenamefont {Tomi}, \citenamefont {Wengler}, \citenamefont {Wiesner},\ and\
  \citenamefont {Hakonen}}]{Fay:2011}%
  \BibitemOpen
  \bibfield  {author} {\bibinfo {author} {\bibfnamefont {A.}~\bibnamefont
  {Fay}}, \bibinfo {author} {\bibfnamefont {R.}~\bibnamefont {Danneau}},
  \bibinfo {author} {\bibfnamefont {J.~K.}\ \bibnamefont {Viljas}}, \bibinfo
  {author} {\bibfnamefont {F.}~\bibnamefont {Wu}}, \bibinfo {author}
  {\bibfnamefont {M.~Y.}\ \bibnamefont {Tomi}}, \bibinfo {author}
  {\bibfnamefont {J.}~\bibnamefont {Wengler}}, \bibinfo {author} {\bibfnamefont
  {M.}~\bibnamefont {Wiesner}}, \ and\ \bibinfo {author} {\bibfnamefont
  {P.~J.}\ \bibnamefont {Hakonen}},\ }\href {\doibase
  10.1103/PhysRevB.84.245427} {\bibfield  {journal} {\bibinfo  {journal} {Phys.
  Rev. B}\ }\textbf {\bibinfo {volume} {84}},\ \bibinfo {pages} {245427}
  (\bibinfo {year} {2011})}\BibitemShut {NoStop}%
\bibitem [{\citenamefont {Kumar}\ \emph {et~al.}(2012)\citenamefont {Kumar},
  \citenamefont {Avriller}, \citenamefont {Yeyati},\ and\ \citenamefont {van
  Ruitenbeek}}]{Kumar:2012}%
  \BibitemOpen
  \bibfield  {author} {\bibinfo {author} {\bibfnamefont {M.}~\bibnamefont
  {Kumar}}, \bibinfo {author} {\bibfnamefont {R.}~\bibnamefont {Avriller}},
  \bibinfo {author} {\bibfnamefont {A.~L.}\ \bibnamefont {Yeyati}}, \ and\
  \bibinfo {author} {\bibfnamefont {J.~M.}\ \bibnamefont {van Ruitenbeek}},\
  }\href {\doibase 10.1103/PhysRevLett.108.146602} {\bibfield  {journal}
  {\bibinfo  {journal} {Phys. Rev. Lett.}\ }\textbf {\bibinfo {volume} {108}},\
  \bibinfo {pages} {146602} (\bibinfo {year} {2012})}\BibitemShut {NoStop}%
\bibitem [{\citenamefont {Weissman}(1988)}]{Weissman:1988}%
  \BibitemOpen
  \bibfield  {author} {\bibinfo {author} {\bibfnamefont {M.~B.}\ \bibnamefont
  {Weissman}},\ }\href {\doibase 10.1103/RevModPhys.60.537} {\bibfield
  {journal} {\bibinfo  {journal} {Rev. Mod. Phys.}\ }\textbf {\bibinfo {volume}
  {60}},\ \bibinfo {pages} {537} (\bibinfo {year} {1988})}\BibitemShut
  {NoStop}%
\bibitem [{\citenamefont {Wu}\ \emph {et~al.}(2008)\citenamefont {Wu},
  \citenamefont {Wu}, \citenamefont {Oberholzer}, \citenamefont {Steinacher},
  \citenamefont {Calame},\ and\ \citenamefont {Sch\"onenberger}}]{Wu:2008}%
  \BibitemOpen
  \bibfield  {author} {\bibinfo {author} {\bibfnamefont {Z.~M.}~\bibnamefont
  {Wu}}, \bibinfo {author} {\bibfnamefont {S.~M.}~\bibnamefont {Wu}}, \bibinfo
  {author} {\bibfnamefont {S.}~\bibnamefont {Oberholzer}}, \bibinfo {author}
  {\bibfnamefont {M.}~\bibnamefont {Steinacher}}, \bibinfo {author}
  {\bibfnamefont {M.}~\bibnamefont {Calame}}, \ and\ \bibinfo {author}
  {\bibfnamefont {C.}~\bibnamefont {Sch\"onenberger}},\ }\href {\doibase
  10.1103/PhysRevB.78.235421} {\bibfield  {journal} {\bibinfo  {journal} {Phys.
  Rev. B}\ }\textbf {\bibinfo {volume} {78}},\ \bibinfo {pages} {235421}
  (\bibinfo {year} {2008})}\BibitemShut {NoStop}%
\bibitem [{\citenamefont {Xu}\ and\ \citenamefont {Tao}(2003)}]{Xu:2003}%
  \BibitemOpen
  \bibfield  {author} {\bibinfo {author} {\bibfnamefont {B.}~\bibnamefont
  {Xu}}\ and\ \bibinfo {author} {\bibfnamefont {N.~J.}\ \bibnamefont {Tao}},\
  }\href {\doibase 10.1126/science.1087481} {\bibfield  {journal} {\bibinfo
  {journal} {Science}\ }\textbf {\bibinfo {volume} {301}},\ \bibinfo {pages}
  {1221} (\bibinfo {year} {2003})}.\BibitemShut
  {NoStop}%
\bibitem [{\citenamefont {Venkataraman}\ \emph {et~al.}(2006)\citenamefont
  {Venkataraman}, \citenamefont {Klare}, \citenamefont {Tam}, \citenamefont
  {Nuckolls}, \citenamefont {Hybertsen},\ and\ \citenamefont
  {Steigerwald}}]{Venkataraman:2006}%
  \BibitemOpen
  \bibfield  {author} {\bibinfo {author} {\bibfnamefont {L.}~\bibnamefont
  {Venkataraman}}, \bibinfo {author} {\bibfnamefont {J.~E.}\ \bibnamefont
  {Klare}}, \bibinfo {author} {\bibfnamefont {I.~W.}\ \bibnamefont {Tam}},
  \bibinfo {author} {\bibfnamefont {C.}~\bibnamefont {Nuckolls}}, \bibinfo
  {author} {\bibfnamefont {M.~S.}\ \bibnamefont {Hybertsen}}, \ and\ \bibinfo
  {author} {\bibfnamefont {M.~L.}\ \bibnamefont {Steigerwald}},\ } {\bibfield  {journal} {\bibinfo  {journal}
  {Nano Letters}\ }\textbf {\bibinfo {volume} {6}},\ \bibinfo {pages} {458}
  (\bibinfo {year} {2006})}\BibitemShut {NoStop}%
\bibitem [{\citenamefont {Agra{\"i}t}\ \emph {et~al.}(2003)\citenamefont
  {Agra{\"i}t}, \citenamefont {Yeyati},\ and\ \citenamefont {van
  Ruitenbeek}}]{Agrait:2003}%
  \BibitemOpen
  \bibfield  {author} {\bibinfo {author} {\bibfnamefont {N.}~\bibnamefont
  {Agra{\"i}t}}, \bibinfo {author} {\bibfnamefont {A.~L.}\ \bibnamefont
  {Yeyati}}, \ and\ \bibinfo {author} {\bibfnamefont {J.~M.}\ \bibnamefont {van
  Ruitenbeek}},\ }\href {\doibase 10.1016/S0370-1573(02)00633-6} {\bibfield
  {journal} {\bibinfo  {journal} {Physics Reports}\ }\textbf {\bibinfo {volume}
  {377}},\ \bibinfo {pages} {81 } (\bibinfo {year} {2003})}\BibitemShut
  {NoStop}%
\bibitem [{\citenamefont {Yanson}\ \emph {et~al.}(2005)\citenamefont {Yanson},
  \citenamefont {Shklyarevskii}, \citenamefont {Csonka}, \citenamefont {van
  Kempen}, \citenamefont {Speller}, \citenamefont {Yanson},\ and\ \citenamefont
  {van Ruitenbeek}}]{Yanson:2005}%
  \BibitemOpen
  \bibfield  {author} {\bibinfo {author} {\bibfnamefont {I.~K.}\ \bibnamefont
  {Yanson}}, \bibinfo {author} {\bibfnamefont {O.~I.}\ \bibnamefont
  {Shklyarevskii}}, \bibinfo {author} {\bibfnamefont {S.}~\bibnamefont
  {Csonka}}, \bibinfo {author} {\bibfnamefont {H.}~\bibnamefont {van Kempen}},
  \bibinfo {author} {\bibfnamefont {S.}~\bibnamefont {Speller}}, \bibinfo
  {author} {\bibfnamefont {A.~I.}\ \bibnamefont {Yanson}}, \ and\ \bibinfo
  {author} {\bibfnamefont {J.~M.}\ \bibnamefont {van Ruitenbeek}},\ }\href
  {\doibase 10.1103/PhysRevLett.95.256806} {\bibfield  {journal} {\bibinfo
  {journal} {Phys. Rev. Lett.}\ }\textbf {\bibinfo {volume} {95}},\ \bibinfo
  {pages} {256806} (\bibinfo {year} {2005})}\BibitemShut {NoStop}%
\bibitem [{\citenamefont {Scheer}\ \emph {et~al.}(2001)\citenamefont {Scheer},
  \citenamefont {Belzig}, \citenamefont {Naveh}, \citenamefont {Devoret},
  \citenamefont {Esteve},\ and\ \citenamefont {Urbina}}]{Scheer:2001}%
  \BibitemOpen
  \bibfield  {author} {\bibinfo {author} {\bibfnamefont {E.}~\bibnamefont
  {Scheer}}, \bibinfo {author} {\bibfnamefont {W.}~\bibnamefont {Belzig}},
  \bibinfo {author} {\bibfnamefont {Y.}~\bibnamefont {Naveh}}, \bibinfo
  {author} {\bibfnamefont {M.~H.}\ \bibnamefont {Devoret}}, \bibinfo {author}
  {\bibfnamefont {D.}~\bibnamefont {Esteve}}, \ and\ \bibinfo {author}
  {\bibfnamefont {C.}~\bibnamefont {Urbina}},\ }\href {\doibase
  10.1103/PhysRevLett.86.284} {\bibfield  {journal} {\bibinfo  {journal} {Phys.
  Rev. Lett.}\ }\textbf {\bibinfo {volume} {86}},\ \bibinfo {pages} {284}
  (\bibinfo {year} {2001})}\BibitemShut {NoStop}%
\bibitem [{\citenamefont {Nagaev}(1995)}]{Nagaev:1995}%
  \BibitemOpen
  \bibfield  {author} {\bibinfo {author} {\bibfnamefont {K.~E.}\ \bibnamefont
  {Nagaev}},\ }\href {\doibase 10.1103/PhysRevB.52.4740} {\bibfield  {journal}
  {\bibinfo  {journal} {Phys. Rev. B}\ }\textbf {\bibinfo {volume} {52}},\
  \bibinfo {pages} {4740} (\bibinfo {year} {1995})}\BibitemShut {NoStop}%
\bibitem [{\citenamefont {Riquelme}\ \emph {et~al.}(2005)\citenamefont
  {Riquelme}, \citenamefont {de~la Vega}, \citenamefont {Yeyati}, \citenamefont
  {Agra{\"i}t}, \citenamefont {Martin-Rodero},\ and\ \citenamefont
  {Rubio-Bollinger}}]{Riquelme:2005}%
  \BibitemOpen
  \bibfield  {author} {\bibinfo {author} {\bibfnamefont {J.~J.}\ \bibnamefont
  {Riquelme}}, \bibinfo {author} {\bibfnamefont {L.}~\bibnamefont {de~la
  Vega}}, \bibinfo {author} {\bibfnamefont {A.~L.}\ \bibnamefont {Yeyati}},
  \bibinfo {author} {\bibfnamefont {N.}~\bibnamefont {Agra{\"i}t}}, \bibinfo
  {author} {\bibfnamefont {A.}~\bibnamefont {Martin-Rodero}}, \ and\ \bibinfo
  {author} {\bibfnamefont {G.}~\bibnamefont {Rubio-Bollinger}},\ } {\bibfield  {journal}
  {\bibinfo  {journal} {Europhys. Lett.}\ }\textbf {\bibinfo {volume} {70}},\
  \bibinfo {pages} {663} (\bibinfo {year} {2005})}\BibitemShut {NoStop}%
\bibitem [{\citenamefont {Ludoph}\ and\ \citenamefont
  {van Ruitenbeek}(2000)}]{Ludoph:2000}%
  \BibitemOpen
  \bibfield  {author} {\bibinfo {author} {\bibfnamefont {B.}~\bibnamefont
  {Ludoph}}\ and\ \bibinfo {author} {\bibfnamefont {J.~M.}\ \bibnamefont
  {van Ruitenbeek}},\ }\href {\doibase 10.1103/PhysRevB.61.2273} {\bibfield
  {journal} {\bibinfo  {journal} {Phys. Rev. B}\ }\textbf {\bibinfo {volume}
  {61}},\ \bibinfo {pages} {2273} (\bibinfo {year} {2000})}\BibitemShut
  {NoStop}%
\bibitem [{\citenamefont {Scheer}\ \emph {et~al.}(1998)\citenamefont {Scheer},
  \citenamefont {Agra{\"i}t}, \citenamefont {Cuevas}, \citenamefont {Yeyati},
  \citenamefont {Ludoph}, \citenamefont {Mart{'i}n-Rodero}, \citenamefont
  {Bollinger}, \citenamefont {van Ruitenbeek},\ and\ \citenamefont
  {Urbina}}]{Scheer:1998}%
  \BibitemOpen
  \bibfield  {author} {\bibinfo {author} {\bibfnamefont {E.}~\bibnamefont
  {Scheer}}, \bibinfo {author} {\bibfnamefont {N.}~\bibnamefont {Agra{\"i}t}},
  \bibinfo {author} {\bibfnamefont {J.~C.}\ \bibnamefont {Cuevas}}, \bibinfo
  {author} {\bibfnamefont {A.~L.}\ \bibnamefont {Yeyati}}, \bibinfo {author}
  {\bibfnamefont {B.}~\bibnamefont {Ludoph}}, \bibinfo {author} {\bibfnamefont
  {A.}~\bibnamefont {Mart{'i}n-Rodero}}, \bibinfo {author} {\bibfnamefont
  {G.~R.}\ \bibnamefont {Bollinger}}, \bibinfo {author} {\bibfnamefont {J.~M.}\
  \bibnamefont {van Ruitenbeek}}, \ and\ \bibinfo {author} {\bibfnamefont
  {C.}~\bibnamefont {Urbina}},\ }\href@noop {} {\bibfield  {journal} {\bibinfo
  {journal} {Nature}\ }\textbf {\bibinfo {volume} {394}},\ \bibinfo {pages}
  {154} (\bibinfo {year} {1998})}\BibitemShut {NoStop}%
\bibitem [{\citenamefont {Makk}\ \emph {et~al.}(2012)\citenamefont {Makk},
  \citenamefont {Tomaszewski}, \citenamefont {Martinek}, \citenamefont
  {Balogh}, \citenamefont {Csonka}, \citenamefont {Wawrzyniak}, \citenamefont
  {Frei}, \citenamefont {Venkataraman},\ and\ \citenamefont
  {Halbritter}}]{Makk:2012}%
  \BibitemOpen
  \bibfield  {author} {\bibinfo {author} {\bibfnamefont {P.}~\bibnamefont
  {Makk}}, \bibinfo {author} {\bibfnamefont {D.}~\bibnamefont {Tomaszewski}},
  \bibinfo {author} {\bibfnamefont {J.}~\bibnamefont {Martinek}}, \bibinfo
  {author} {\bibfnamefont {Z.}~\bibnamefont {Balogh}}, \bibinfo {author}
  {\bibfnamefont {S.}~\bibnamefont {Csonka}}, \bibinfo {author} {\bibfnamefont
  {M.}~\bibnamefont {Wawrzyniak}}, \bibinfo {author} {\bibfnamefont
  {M.}~\bibnamefont {Frei}}, \bibinfo {author} {\bibfnamefont {L.}~\bibnamefont
  {Venkataraman}}, \ and\ \bibinfo {author} {\bibfnamefont {A.}~\bibnamefont
  {Halbritter}},\ }\href {\doibase 10.1021/nn300440f} {\bibfield  {journal}
  {\bibinfo  {journal} {ACS Nano}\ }\textbf {\bibinfo {volume} {6}},\ \bibinfo
  {pages} {3411} (\bibinfo {year} {2012})}.\BibitemShut {NoStop}%

\end{thebibliography}


%

\end {document}